\documentclass[sigplan,10pt]{acmart}
\renewcommand\footnotetextcopyrightpermission[1]{}
\usepackage{amsmath}
\usepackage{graphicx}
\usepackage{xcolor}
\usepackage{cleveref}
\usepackage{subcaption}
\usepackage{footnote}
\usepackage{caption}
\makesavenoteenv{figure}
\crefname{figure}{Figure}{Figures}
\Crefname{figure}{Figure}{Figures}
\usepackage{makecell}

\usepackage{booktabs}
\usepackage{tabularx}
\usepackage{array}
\newcolumntype{P}[1]{>{\raggedright\arraybackslash}p{#1}}

\usepackage{algorithm}
\usepackage[noend]{algpseudocode}

\newcommand{\jpara}[1]{%
  \vspace{3pt}%
  \noindent\textbf{{#1}}
}

\usepackage{enumitem}

\newenvironment{myitemize}
{\begin{itemize}[
    leftmargin=1.75em,
    topsep=1pt,
    partopsep=1pt,
    itemsep=1pt,
    parsep=1pt
]}
{\end{itemize}}

\title{The Energy Cost of Execution-Idle in GPU Clusters}
\author{Yiran Lei, Jared Fernandez, Vasilis Kypriotis, Dimitrios Skarlatos,\\Emma Strubell, Justine Sherry, Daniel Vosler}

\begin{abstract}
GPUs are becoming a major contributor to data center power, yet unlike CPUs, they can remain at high power even when visible activity is near zero.
We call this state \emph{execution-idle}.
Using per-second telemetry from a large academic AI cluster, we characterize execution-idle as a recurring low-activity yet high-power state in real deployments.
Across diverse workloads and multiple GPU generations, it accounts for 19.7\% of in-execution time and 10.7\% of energy.
This suggests a need to both reduce the cost of execution-idle and reduce exposure to it.
We therefore build two prototypes: one uses automatic downscaling during execution-idle, and the other uses load imbalance to reduce exposure, both with performance trade-offs.
These findings suggest that future GPU systems should treat execution-idle as a first-class operating state.
\end{abstract}

\begin{document}
\maketitle
\pagestyle{plain}

\section{Introduction}
\label{sec:intro}

AI’s growing power demand is becoming a significant environmental concern.
Data centers consume roughly 4--5\% of U.S. electricity and are projected to reach as much as 17\% by 2030~\cite{epri2026powering_intelligence,shehabi2024datacenterenergy,mckinsey2024ai_power}.
GPUs are a major driver of this trend: they account for about 60\% of power in multi-GPU servers~\cite{characterize_llm_power} and roughly 41\% of total power in AI clusters~\cite{emberson2025gpus_power_ai_datacenters}.
Yet current understanding of AI energy use is still dominated by coarse aggregate metrics, such as total GPU-hours and Thermal Design Power (TDP)~\cite{bloom, llama3,inference_measurement1}.
While these metrics convey the overall scale of energy consumption, they obscure how GPU power evolves during execution.
As we show in this paper, a fine grained view of runtime behavior reveals where GPU energy is spent productively and where it is not.

We observe that a GPU can continue drawing substantial power even when a live job shows little compute, memory, or communication activity.
This behavior is counterintuitive: one might expect an underutilized GPU to consume relatively little energy, but our measurements show otherwise.
In state-of-the-art serving traces, such intervals can account for up to 65\% of total energy use.
We call this state \emph{execution-idle}: intervals during execution in which the GPU remains allocated and a program remains loaded, yet visible activity is near zero.
This differs from truly idle periods, in which the GPU is unused and returns to baseline power.

\begin{figure}
    \centering
    \includegraphics[width=0.8\linewidth]{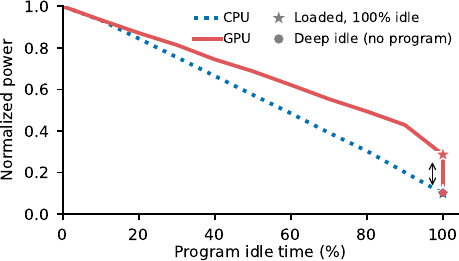}
    \caption{CPU power falls with idle time, but GPU power remains elevated even when a loaded program is fully idle.}
    \label{fig:gpu_versus_cpu}
\end{figure}

Execution-idle states are easy to overlook for two reasons.
First, CPU-based energy intuition~\cite{cpu_energy1,cpu_energy2} does not carry over cleanly to GPUs.
As \cref{fig:gpu_versus_cpu} shows\footnote{We run a matrix multiplication benchmark with a configurable pause fraction on an Intel Xeon 6226R and an NVIDIA L40S. CPU power includes package and DRAM power measured with Intel VTune; GPU power is total device power reported by \texttt{nvidia-smi}. The figure excludes other server components such as the motherboard and storage.}, CPU power typically tracks inactivity more closely, whereas GPU power can remain elevated even when a loaded program is fully idle.
Second, with few exceptions~\cite{amd_profiling,characterize_llm_power}, prior work analyzes GPU energy at a coarse granularity, for example through end-to-end summaries or comparisons across different load levels~\cite{energy_rps1,energy_tokennum_batch,energy_model_task_gpu,energy_batch_models,energy_batchsize_limit,engine_benchmark,energy_frequency1,energy_rps2,single_node_measurement,ml_energy,nvidia_energy_efficiency_blog_2024,energy_frequency2,energy_frequency3,envpipe,bloom,characterize_llm_power,inference_measurement1,serving_frequency,energy_tp_request_length_task_idle,blackwell_power_profiles,blackwell_power_profiles_paper}.
What remains missing is a fine-grained characterization of how GPU power and visible activity co-evolve during execution in real deployments.

In this paper, we study execution-idle as a recurring operating regime in modern GPU systems.
To do so, we build a passive profiling system that collects per-second GPU power and utilization telemetry.
We deploy it for 31 days on a large academic AI cluster. To validate that execution-idle states can occur outside of the university setting, we complement our data from academic, experimental workloads with replays of production, industrial serving workloads.
This combined view lets us define execution-idle explicitly, quantify its prevalence and energy cost for a range of settings, and examine causes for why it arises in these settings.
All measurements are collected under the cluster’s standard performance-oriented production configuration~\cite{hpe_hpc_profile}, without power caps or manual tuning.

Our measurements show that execution-idle is a non-trivial component of GPU energy use across workloads, platforms, and serving environments.
\begin{myitemize}
\item \textbf{Execution-idle appears across all six GPU platforms we study.}
We observe sustained execution-idle intervals in both training and inference jobs across six GPU types, including NVIDIA B200~\cite{b200}, showing that this phenomenon is not confined to a particular workload or hardware generation.

\item \textbf{Execution-idle is especially costly for serving workloads.}
Bursty arrivals create loaded-but-inactive gaps, making execution-idle a substantial source of serving energy.
It accounts for 48\% of energy in long-lived, academic serving workloads and 7--65\% across five replays of industry-derived traces from OpenAI~\cite{BurstGPT}, Qwen~\cite{qwenbailian}, and Azure~\cite{energy_tp_request_length_task_idle}.
This is particularly important because serving is projected to account for up to 70\% of energy use in industry clusters~\cite{serving_cost1,serving_cost2}.

\item \textbf{Execution-idle also significantly contributes energy cost to non-serving workloads we observed.}
Our academic workloads also included training and batched inference workloads, which spent 6\% and 7\% of total energy consumption in execution-idle states respectively; with 13\% and 12\% of total execution time in execution-idle states.

\item \textbf{Many execution-idle intervals are associated with I/O bottlenecks.}
By examining the activity immediately preceding execution-idle intervals, we find that they often follow PCIe transfers (48\% of cases), network-backed I/O (17\%), NVLink communication (2\%), and other events, suggesting that execution-idle likely arises from software bottlenecks waiting on I/O.

\end{myitemize}

Together, our measurements show that relatively brief stalls can become a meaningful source of energy cost. 
From the perspective of hardware researchers, the cost of execution-idle states points to a need for continued research towards power-proportional GPU design.
From the perspective of software systems researchers, we must make due with the hardware we have, with mechanisms to either lower the cost or execution-idle periods or reduce exposure to execution-idle periods in the first place.

Unfortunately, na\"{i}ve approaches to cap GPU energy during execution-idle states (reducing cost) or to increase utilization (reducing the time spent in execution idle states) perhaps predictably save energy, but also increase response latency.
To explore whether energy waste due to execution-idle states has an easy fix, we implemented two simple prototypes.
One applies automated fine-grained downscaling to reduce power during execution-idle.
The other applies deliberate load imbalance in serving to consolidate work onto fewer GPUs, allowing the remaining GPUs to stay idle for longer intervals and avoid execution-idle altogether.
Both reduce energy, but both also substantially increase latency, showing that execution-idle is actionable but not free to manage with na\"{i}ve techniques.

Overall, we argue that execution-idle should be a first-class concern for future energy-efficient GPU systems.
Hardware should better approach power proportionality during execution-idle, while software should either keep GPUs more fully utilized or coordinate explicitly with power-management mechanisms across layers.
Achieving this in practice, however, will require future research to more effectively navigate fundamental energy--performance trade-offs.

\jpara{Paper roadmap.}
The rest of the paper is organized as follows. \S\ref{sec:datasets} presents our characterization methodology, including the definition of execution-idle and the measurement scope. \S\ref{sec:execution_idle} provides initial observations, and \S\ref{sec:findings} reports the main findings on its prevalence and energy cost, especially for serving workloads. \S\ref{sec:implication} discusses implications, \S\ref{sec:future} outlines broader system support for future GPU system design, and \S\ref{sec:related} reviews related work.
\section{Datasets \& Characterization Methodology}
\label{sec:datasets}

\begin{table*}[t]
\centering
\footnotesize
\setlength{\tabcolsep}{5pt}
\renewcommand{\arraystretch}{1.12}
\caption{Representative signals collected by our passive profiling pipeline. These signals cover GPU power, activity, clocks, communication, host activity, timing, and job metadata.}
\label{tbl:telemetry}
\begin{tabularx}{\textwidth}{P{1.7cm} P{3.5cm} P{0.8cm} P{1.7cm} X}
\toprule
\textbf{Domain} & \textbf{Metric} & \textbf{Unit} & \textbf{Source} & \textbf{Description} \\
\midrule

GPU identity
& \texttt{hostname}, \texttt{gpu\_id}, \texttt{gpu\_name}
& --
& Slurm, NVML
& Identifies the host, GPU instance, and GPU model for each sampled record. \\
\midrule

GPU power
& \texttt{power}
& W
& NVML
& GPU board power used for energy accounting. \\
\midrule

GPU activity
& \texttt{sm}, \texttt{tensor}, \texttt{dram}, \texttt{fp16}, \texttt{fp32}, \texttt{fp64}
& \%
& DCGM
& Compute and memory activity signals used to characterize whether the GPU is actively doing work. Some counters are unavailable on some GPU types. \\
\midrule

GPU clocks
& \texttt{sm\_clk}, \texttt{mem\_clk}
& MHz
& NVML
& Runtime GPU clock frequencies used to study frequency behavior under different execution states. \\
\midrule

GPU communication
& \texttt{pcie\_tx}, \texttt{pcie\_rx}, \texttt{nvlink\_tx}, \texttt{nvlink\_rx}
& MB/s
& NVML, \texttt{nvidia-smi}
& Device-side communication signals used to track PCIe and NVLink traffic. NVLink counters are unavailable on some GPU types. \\
\midrule

Host activity
& \texttt{cpu\_util}, \texttt{host\_mem\_util}
& \%
& \texttt{psutil}
& Host-side CPU and memory utilization used to characterize surrounding system activity. \\
\midrule

Network activity
& \texttt{nic\_tx}, \texttt{nic\_rx}
& MB/s
& OS counters
& Per-interface network throughput used to capture external data movement and network activity. \\
\midrule

Timing
& \texttt{timestamp}
& s
& Profiler
& Per-sample timestamp used to align telemetry across sources and over time. \\
\midrule

Job metadata
& \texttt{job\_id}, \texttt{job\_name}
& --
& Slurm
& Scheduler metadata used to associate telemetry with jobs and support job-level analysis. \\
\bottomrule
\end{tabularx}
\end{table*}

This section describes the dataset and methodology we use to characterize execution-idle in GPU workloads.
We first describe the measurement environment and passive telemetry pipeline (\S\ref{subsec:measurement_setting}), then define a taxonomy of GPU states (\S\ref{subsec:taxonomy}) and clarify the scope and limits of what these measurements can reveal (\S\ref{subsec:scope_limitation}).

\subsection{Measurement Setting and Data Collection}
\label{subsec:measurement_setting}

\jpara{Cluster and study window.}
Our study uses passive telemetry from a large academic AI cluster running diverse GPU workloads, including training, batch inference, and online serving.
Jobs are managed by Slurm~\cite{slurm} and receive exclusive whole-GPU allocations, without hardware partitioning such as MIG~\cite{mig}.
All measurements are collected under the cluster's standard production configuration~\cite{hpe_hpc_profile}: nodes use performance-oriented BIOS profiles, and GPUs operate under vendor-managed DVFS without fixed application clocks or additional power caps.
Hence, our data characterizes how execution-idle appears under the default power behavior, rather than any manually tuned or customized settings.\footnote{Recent Blackwell systems introduce data-center power profiles such as Max-Q, which improve efficiency through coarse-grained control across workload classes~\cite{blackwell_power_profiles,blackwell_power_profiles_paper}. These mechanisms are orthogonal to our study, since they do not perform fine-grained temporal adjustments to loaded-but-inactive intervals within executions.}

Our measurement includes 756 NVIDIA GPUs spanning multiple generations: 200 A6000~\cite{gpu_a6000}, 52 RTX 6000 Ada~\cite{gpu_6000ada}, 408 L40(S)~\cite{l40s}, 64 A100~\cite{a100}, 24 H100~\cite{h100}, and 8 B200~\cite{b200}.
Our dataset spans 31 days, from February 4, 2026 to March 7, 2026, and contains 162~GB of telemetry.
Additional cluster details are listed in supplementary material.

\jpara{Passive telemetry pipeline.}
We collect per-second GPU-, host-, and job-level telemetry from NVML~\cite{nvidia_nvml}, DCGM~\cite{nvidia_dcgm}, OS counters, psutil~\cite{rodola_psutil}, nvidia-smi~\cite{nvidia_smi_docs}, and Slurm~\cite{slurm}.
These signals cover GPU power, activity, clocks, communication, host activity, timing, and job metadata; \Cref{tbl:telemetry} summarizes representative fields.

Telemetry is collected passively on each compute node and then aligned with scheduler records using timestamps and GPU allocation metadata, allowing each GPU-second sample to be attributed to a job.
Each retained sample therefore represents one second of behavior on one allocated GPU for one job.
We discard malformed records and samples that cannot be attributed reliably.

\jpara{Profiling overhead.}
The profiling pipeline adds negligible overhead to compute nodes.
According to \texttt{pidstat}~\cite{sysstat_pidstat}, the profiling process remains below 0.05\% CPU and uses roughly 400\,MB of memory, about 0.08\% of the nodes with more than 500\,GB of host memory.
Compressed telemetry logs require only 20--100\,MB per server per day, negligible relative to the more than 5\,TB of local storage available on each server.

\jpara{User Privacy and Ethics.} 
Under our institution's policies, this research is considered exempt from review as our research does not concern human subjects. 
Nonetheless, we are sensitive to privacy concerns from cluster users. 
Hence, we spoke with PIs and cluster stakeholders to inform them of our study; we also supported an `opt-out' flag for users to prevent data about their jobs from being captured.
Finally, the dataset is anonymized and stripped of sensitive information that could link it to any particular researcher.

\subsection{Defining Execution-Idle, Active, \& Deep Idle States}
\label{subsec:taxonomy}

\jpara{Definitions} The \emph{execution-idle state} consists of intervals during job execution in which the GPU remains allocated and the program remains resident, yet visible activity is near zero.
We classify an interval as execution-idle when all available compute- and memory-related signals---including SM, tensor-core, other per-precision accelerator activity (e.g., \texttt{fp32}) when available, and DRAM activity---remain below 5\%, and all available communication signals---including PCIe and NVLink traffic when available---remain below 1\,GB/s ($\approx$3\% of PCIe 4.0 $\times$ 16 bandwidth).
These conditions must hold simultaneously.
If a signal is unavailable on a given GPU type, we omit it from the rule rather than treating it as violated.

Alongside \emph{execution-idle}, we distinguish two other GPU states: \emph{deep idle}, in which no program is resident and the GPU remains at baseline power, and \emph{active execution}, in which a program is resident and observed activity exceeds the execution-idle threshold.
Thus, \emph{execution-idle} and \emph{active execution} are both in-program states, whereas \emph{deep idle} is not.
The three states are mutually exclusive and collectively exhaustive.
The key distinction is that, unlike \emph{deep idle}, \emph{execution-idle} occurs within a live job: a program remains resident, yet visible activity is low, and power can stay substantially above baseline despite little or no useful work.

We quantify time by summing samples assigned to each state.
We quantify energy using NVML-reported board power: integrating power over all samples yields total GPU energy, while integrating only over execution-idle samples yields execution-idle energy.
The ratio of these two quantities gives the fraction of GPU energy spent in execution-idle.

\jpara{Conservative quantification of execution-idle intervals.}
Modern GPUs use dynamic voltage and frequency scaling (DVFS)~\cite{dvfs_survey} to adjust operating frequency in response to workload conditions, which in turn affects both device power and performance.
Prior work suggests that modern GPUs may take roughly 1--500\,ms to adjust frequency~\cite{frequency_scaling_time}.
Very short pauses may therefore be too brief for frequency to fall, recover, and manifest as a distinct operating regime.

To avoid counting such transient gaps as execution-idle when quantifying the cluster workloads, we adopt a conservative duration threshold of 5\,s: an interval is counted as execution-idle only if it satisfies the low-activity conditions continuously for at least 5\,s.
This threshold is long enough to exclude brief pauses that existing DVFS mechanisms are intended to absorb, while still capturing sustained low-activity intervals that are long enough to matter for measurement and, potentially, system response.
As a result, our methodology likely \emph{underestimates}, rather than overstates, how often execution-idle occurs, and our main conclusions are not sensitive to this threshold (\S\ref{subsec:robustness}).

\subsection{Scope and Limitations}
\label{subsec:scope_limitation}

\jpara{Primary AI research cluster analysis.}
Our dataset comes from an academic AI cluster rather than a specialized production deployment, so its aggregate utilization is not meant to represent all industry GPU fleets. 
Compared with many production settings, academic clusters expose a broader mix of workload classes and user behaviors, including research workloads, short jobs, debugging runs, and interactive sessions. 
We therefore focus the later analyses on long-running, non-debug jobs, which better capture sustained training, batch inference, and serving behavior and more closely reflect where GPU energy is spent. 
This breadth is useful for characterization, since it reveals how execution-idle varies across workload classes within one environment. 
At the same time, many production GPU deployments are more specialized—for example, around training~\cite{xai_colossus} or latency-sensitive serving---so bottlenecks that are diluted in a mixed academic trace may be even sharper in those settings. We examine that next with serving-oriented workloads.

For broad workload-level comparisons, we group academic jobs using keyword-based rules over job metadata; yielding the following groups of jobs: academic serving, academic batch inference, academic training, and academic others.
These labels are intended only for coarse grouping rather than precise semantic classification.

\jpara{Complementary industry-style replay.}
To corroborate that we might expect to see execution-idle states outside of academic workloads, we also study industry-style GPU usage using open-source serving systems driven by public traces derived from OpenAI~\cite{BurstGPT}, Qwen~\cite{qwenbailian}, and Azure~\cite{energy_tp_request_length_task_idle}.
These traces provide request arrival times together with input and output token lengths.
Following state-of-the-art trace-driven serving evaluations~\cite{energy_tp_request_length_task_idle}, we synthesize requests from the traced token lengths and serve an open-source Llama-13B model~\cite{llama} on L40S GPUs using vLLM~\cite{vllm}.
Because the original traces were collected from deployments with larger fixed GPU pools than our testbed, we downscale each trace to a smaller but still fixed pool while preserving burstiness, and replay the resulting per-GPU streams for 30 minutes.
This setup preserves the fixed-provisioning assumption of the original traced deployments (e.g., 32 GPUs in Qwen~\cite{qwenbailian} and 96 GPUs in Azure~\cite{energy_tp_request_length_task_idle}); autoscaling~\cite{autoscale1,autoscale2,autoscale3} is therefore outside the scope of our replay study.
Because the replay uses one model, one GPU type, and one serving engine, it is not intended to fully characterize the magnitude of execution-idle across serving deployments.
Instead, it serves as a complementary case study to test whether execution-idle also arises under more production-like serving demand.

\jpara{Temporal and observability limits.}
Our passive measurements have two important limits.
First, telemetry is sampled at 1\,Hz, so very short bursts and sub-second transitions may be smoothed or missed.
We therefore focus on sustained behavior rather than microsecond- or millisecond-scale dynamics.
This is a deliberate trade-off: prior work~\cite{amd_profiling} uses finer-grained logging to study short kernels and micro-kernels, whereas our goal is to characterize long-running deployed jobs, where sustained power draw accumulates into meaningful energy cost.
At the scale of 756 GPUs over 31 days, 1\,Hz sampling provides a practical and scalable resolution while preserving the temporal structure needed for cluster-scale energy accounting.

Second, current public vendor telemetry provides only limited component-level power visibility.
NVIDIA interfaces expose device-level GPU power, and some newer datacenter platforms also report module or memory-subsystem power~\cite{nvidia_smi_docs}, but they do not expose a portable breakdown across compute, memory, and interconnect components.
Because such counters are not available consistently across GPU generations in our cluster, we can quantify execution-idle and its device-level energy cost, but cannot attribute that cost uniformly to specific hardware components.

\section{A First Look at Execution-Idle}
\label{sec:execution_idle}

\begin{figure}[t]
    \includegraphics[width=0.95\linewidth]{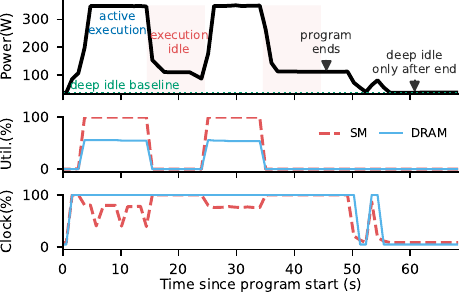}
    \caption{Time-aligned power, SM and DRAM utilization, and normalized frequency for a job on an L40S GPU, illustrating the execution-idle state.}
    \label{fig:exec_idle_example}
\end{figure}

Having defined our methodology, we now present an overview of the prevalence of execution-idle in our AI cluster, how the execution-idle state occurs, the magnitude of execution-idle power use relative to deep idle power use, hypothesize as to why execution-idle states occur, and demonstrate that execution-idle states exist across six generations of GPUs. 
In \S\ref{sec:findings}, we turn to quantifying the overall energy cost of execution-idle states within particular classes of workload.

\begin{figure}[t]
    \centering
    \begin{subfigure}[t]{0.495\linewidth}
        \centering
        \includegraphics[width=\linewidth]{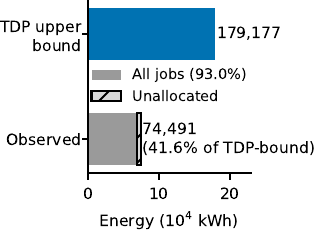}
        \caption{Unallocated/job-attributed energy vs. TDP upper bound}
        \label{fig:cluster_energy_vs_tdp}
    \end{subfigure}
    \begin{subfigure}[t]{0.495\linewidth}
        \centering
        \includegraphics[width=\linewidth]{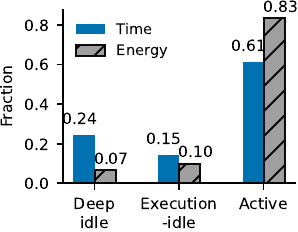}
        \caption{Job-attributed GPU time and energy breakdown}
        \label{fig:time_energy_breakdown}
    \end{subfigure}
    \caption{Cluster-scale GPU energy accounting over the study window. The left panel compares observed GPU energy while the right panel decomposes job-attributed GPU time and energy by regime.}
    \label{fig:cluster_energy_accounting}
\end{figure}

\jpara{Prevalence of execution-idle across the cluster.}
To focus on long-running workloads rather than debug or interactive sessions, we restrict attention to jobs lasting at least two hours (\S\ref{subsec:robustness} evaluates sensitivity to this threshold).
Even after a job is allocated to GPUs, those GPUs may not be used immediately: some jobs begin with CPU- or I/O-heavy setup, such as downloading a large model, leaving the GPU in deep idle.
As a result, job-attributed GPU time and energy still fall into three states: deep idle, execution-idle, and active execution.

\Cref{fig:time_energy_breakdown} shows that deep idle accounts for 24\% of job-attributed GPU time but only 7\% of energy, consistent with the low power draw of hardware with no loaded program.
Execution-idle, by contrast, accounts for 15\% of time and 10\% of energy.
So while execution-idle occupies less time than deep idle, it consumes more energy because the program remains loaded and the GPU stays at elevated power.
\footnote{For one cluster at our university alone, execution-idle states cost an estimated \$944 (at a US average rate of 13.6\textcent~ per kWh), and produced approximately 2.58 – 2.80 metric tons of CO$_2e$ (at a US average rate of 0.82-0.89 lbs of CO$_2e$ per kWh) in a single month.}.
Active execution accounts for the remaining 61\% of time and 83\% of energy, and therefore dominates in-job GPU energy use.

\jpara{How execution-idle appears during a job.}
Figure~\ref{fig:exec_idle_example} illustrates all three GPU states in a measured academic job on an L40S GPU.
During \emph{active execution}, power rises to the expected high-power regime.
During the highlighted \emph{execution-idle} intervals, which last about 10\,s, the job remains resident while visible compute, memory, and PCIe activity all drop to near zero, yet power stays around 110\,W.
Only after the program terminates does the GPU enter \emph{deep idle}, where power drops to the baseline level of roughly 35\,W.
This example shows why execution-idle is a distinct operating state: it occurs within a live job, satisfies our low-activity definition, and still draws far more power than true idle.

\jpara{Why execution-idle power remains high.}
As shown in~\cref{fig:cluster_energy_vs_tdp}, observed GPU energy use over the study is far below the fleet’s rated Thermal Design Power (TDP) upper bound. The cluster’s GPUs consumed 74{,}491\,kWh in total, with 93\% attributable to user jobs and the remaining 7\% to unallocated or deep-idle periods. Overall, this is only 41.6\% of the energy the same GPUs would have consumed if they had run continuously at TDP over the same interval.
This difference reflects the fact that GPUs spend substantial time outside sustained active execution.

\begin{figure}[t]
    \centering
    \includegraphics[width=0.9\linewidth]{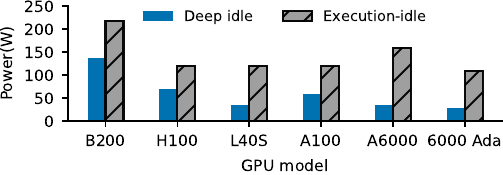}
    \caption{Power in the execution-idle state remains substantially above deep idle across all GPU models in our study.}
    \label{fig:exec_idle_across_gpus}
\end{figure}

Execution-idle highlights why being below TDP is not the same as being energy-efficient. During execution-idle, visible activity drops to near zero, yet \emph{clock frequencies often remain high}, so power stays far above deep idle even though little useful work is being performed.

\begin{figure*}[t]
    \centering
    \includegraphics[width=\linewidth]{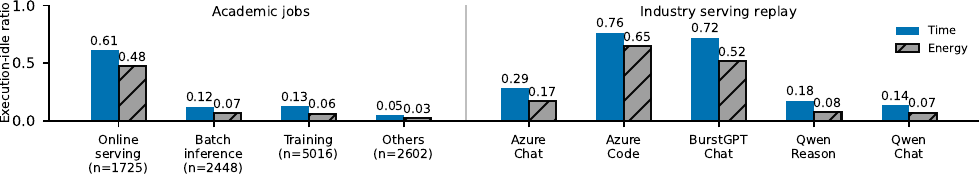}
    \caption{Execution-idle time and energy fractions across academic workload categories and replayed industry serving traces.}
    \label{fig:all_workload_category}
\end{figure*}

We hypothesize that this behavior occurs due to an inability to guarantee fast response latency under truly deep idle states, and a choice to optimize for latency over energy efficiency.
Power managers are tuned to ride through brief within-execution stalls by keeping clocks elevated, preserving responsiveness when work resumes quickly.
That choice becomes costly when low-activity intervals persist (e.g., 10 s).
In our cluster, across 4{,}596{,}723 execution-idle intervals, the median execution-idle duration is 9\,s and the 90th percentile is 44\,s.
At those timescales, elevated frequency is no longer merely a responsiveness mechanism; it becomes a sustained source of energy cost.
Accordingly, execution-idle accounts for 10.7\% of runtime energy across 11{,}791 long-running jobs in the academic dataset.

We further test whether GPU power and frequency eventually downscale during prolonged execution-idle.
Using a controlled experiment that extends execution-idle from 4\,s to 2048\,s, we find that both remain elevated even after 2048\,s.
Prolonged loaded-but-inactive intervals therefore remain power-disproportionate under default GPU behavior.

\jpara{Execution-idle across GPU generations.}
Figure~\ref{fig:exec_idle_across_gpus} shows that execution-idle appears across all GPU types in our study.
Across every GPU model we measure, execution-idle draws substantially more power than deep idle, although the size of this gap varies across architectures and generations.
This variation likely reflects hardware differences, but is difficult to attribute more precisely because comparable power breakdowns are not uniformly available across GPU generations.
Still, the qualitative pattern is consistent: execution-idle appears across modern GPUs, including state-of-the-art NVIDIA B200~\cite{b200} GPUs.

\section{Quantifying Execution-Idle Energy Waste}
\label{sec:findings}

Using the dataset and execution-idle definition from \S\ref{sec:datasets}, we answer five questions:

\begin{myitemize}
    \item How does execution-idle vary across workload categories, including academic jobs and replayed industry serving traces? (\S\ref{subsec:heterogeneous_workloads})
    \item How much time and energy does execution-idle consume across jobs? (\S\ref{subsec:execution_idle_cost})
    \item How sensitive is execution-idle to job lengths and inactivity requirements (\S\ref{subsec:robustness})? 
    \item How long do execution-idle intervals last? (\S\ref{subsec:temporal_structure})
    \item What likely causes tend to precede execution-idle? (\S\ref{subsec:preceding_signals})
\end{myitemize}

Together, these analyses show that execution-idle is a meaningful but uneven component of GPU energy use. It accounts for a nontrivial share of cluster-wide GPU energy, is especially pronounced in serving workloads, is heavy-tailed across jobs, often persists long enough to matter, and follows recurring transfer, communication, and post-compute phases.

\jpara{In-execution fractions.}
Before presenting the results in this section, we clarify the denominator used for all reported fractions.
We are interested in the cost of execution-idle relative to the time and energy spent when a program is running on the GPU.
For that reason, we exclude all deep-idle time and energy (where no program is active) from the denominator, including both (1) periods when no job is allocated to the GPU and (2) deep-idle periods within allocated jobs, such as during CPU- or I/O-heavy setup before the GPU is actually used.

The denominator therefore consists only of \emph{execution-idle} and \emph{active execution}.
This lets us ask: once a program is on the GPU, what fraction of execution time and energy is spent idle but still drawing elevated power?
We call these \emph{in-execution} fractions.

\subsection{Variability of Execution-Idle Across Workloads}
\label{subsec:heterogeneous_workloads}

\jpara{Execution-idle states appear in all workload categories.}
Using the workload labels described in \S\ref{sec:datasets}, we group academic jobs into coarse categories and compare their execution-idle time and energy fractions, alongside the same metrics for replayed industry serving traces. \Cref{fig:all_workload_category} shows that execution-idle appears in every category we examine, indicating that it is not confined to a single workload class.

\begin{figure}[t]
    \centering
    \includegraphics[width=0.7\linewidth]{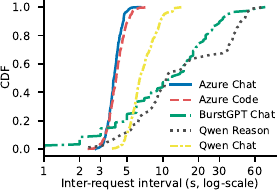}
    \caption{CDF of per-GPU inter-request intervals for replayed industry serving traces.}
    \label{fig:inter-request-interval-cdf}
\end{figure}

\jpara{Execution-idle cost is highest in serving workloads.}
Although execution-idle occurs in all jobs we observed, its magnitude varies sharply across workload categories. 
Among academic jobs, online serving (i.e., handling live latency-sensitive requests) is by far the most exposed regime: GPUs spend 61\% of in-execution time in execution-idle, and 48\% of energy is consumed during those intervals. 
Batch inference (i.e., offline processing over fixed inputs) and training spend much less energy and time in the execution-idle state, at 12--13\% of time and 6--7\% of energy, while the remaining workloads are lower still, at 5\% of time and 3\% of energy. Execution-idle is therefore broadly present, but its cost is highly uneven across workload classes.

\jpara{Serving is especially exposed because GPUs remain resident under bursty demand.}
Serving keeps model state loaded on the GPU to preserve responsiveness, but request arrivals are uneven over time. As a result, GPUs often remain allocated and ready while little or no request work is actively executing, creating long loaded-but-low-activity intervals that still draw substantial power. 
Training and batch inference also exhibit execution-idle, but less persistently. In \S\ref{subsec:preceding_signals}, we explore system conditions that correlate with execution idle-states and find that I/O use often preceeds execution-idle, suggestive of applications blocking on I/O. 

\jpara{Industry serving workloads spent between 14-76\% of time and 7-65\% of energy in execution-idle states.}
To test whether the serving-side effect observed in our academic cluster also appears under industry-style demand, we replay several public serving traces using the method described in \S\ref{sec:datasets}.
Unlike the cluster study, where we impose a conservative 5\,s minimum to avoid counting brief transients in opaque production jobs, the replay setting exposes the request arrival process directly.
We therefore analyze all inter-request low-activity gaps in replay, rather than only those lasting at least 5\,s.
\Cref{fig:inter-request-interval-cdf} shows that per-GPU request streams remain bursty, with the time between consecutive requests often lasting several seconds.
Median inter-request intervals are roughly 4--8\,s across traces, while BurstGPT Chat and Qwen Reason exhibit heavier tails that extend well beyond 10\,s.
These results show that realistic serving traces naturally create frequent loaded-but-low-activity gaps.

Under this replay-specific accounting, replaying the industry serving trace on an L40S GPU, the results in \Cref{fig:all_workload_category} mirror the qualitative pattern seen in the cluster.
Low-activity periods account for 29\%/17\% of time/energy for Azure Chat, 76\%/65\% for Azure Code, 72\%/52\% for BurstGPT Chat, 18\%/8\% for Qwen Reason, and 14\%/7\% for Qwen Chat.
Execution-idle is therefore not unique to the serving jobs observed in our cluster; it also appears under replayed industry demand traces.
Its magnitude varies across traces, and more broadly will depend on the model, serving system, and GPU platform.

\begin{figure}[t]
    \centering
    \includegraphics[width=0.7\linewidth]{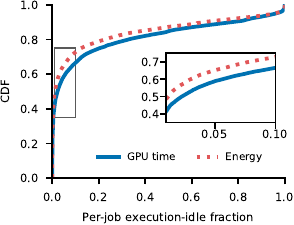}
    \caption{CDF of per-job execution-idle time and energy fractions.}
    \label{fig:per_job_cdf}
\end{figure}

\jpara{Both request spacing and duration shape execution-idle.}
Differences across traces reflect both how long GPUs wait between requests and how long each request occupies the GPU once admitted. Reasoning-heavy requests such as Qwen Reason keep the GPU busy longer, which reduces the fraction of time spent in execution-idle despite relatively long inter-request gaps. By contrast, shorter requests such as Azure Chat, and especially Azure Code, return the GPU to a loaded-but-inactive state more quickly, making second-scale gaps more costly.

Taken together, these results show that execution-idle is a general GPU phenomenon, but one whose cost is especially concentrated in serving. The replayed traces also point to a concrete driver: bursty demand acting on GPUs that remain resident and ready between requests.

\subsection{Distribution of Execution Idle States Per Job}
\label{subsec:execution_idle_cost}

\jpara{The distribution of job time spent in execution-idle is highly right-skewed, with 15.4\% of jobs spending more than half of their time in execution-idle.}
We next examine how execution-idle is distributed across individual jobs. \Cref{fig:per_job_cdf} shows the CDF of per-job execution-idle fractions in both time and energy. A substantial tail of jobs spends a large share of time in this state: 33.4\% of jobs spend more than 10\% of time in execution-idle, 25.2\% spend more than 20\%, and 15.4\% spend more than half of their time in execution-idle.

Some of this tail likely reflects serving-like burstiness, but it is not confined to serving. Among 11,791 long-running jobs, only 1,725 are confirmed serving jobs, or 14.6\% of the population. Therefore, even under the extreme assumption that every serving job falls into the $>$20\% execution-idle tail, at least 10.6\% of all jobs in that tail must come from other workload categories. Execution-idle tail behavior therefore extends beyond serving and affects a broader set of workloads.

\jpara{Highly right-skewed tail also appears in energy distribution.}
Overall, 27.1\% of jobs spend more than 10\% of energy in execution-idle, 21.2\% spend more than 20\%, and 12.8\% spend more than half of energy in this state. Thus, execution-idle is not merely common on average; it is a severe energy cost for a nontrivial tail of jobs.

\subsection{Effects of Inactivity Requirements and Job Length on Execution-Idle.}
\label{subsec:robustness}
We next test whether our main estimates depend strongly on two conservative analysis choices: the long-job cutoff used to focus on sustained workloads and the minimum interval duration used to separate sustained execution-idle from brief transients.

\begin{table}[t]
\centering
\footnotesize
\setlength{\tabcolsep}{5pt}
\renewcommand{\arraystretch}{1.12}
\caption{Execution-idle estimates during in-job execution under alternative thresholds.}
\label{tbl:robustness}
\begin{tabular}{p{2.5cm}cccc}
\toprule
\textbf{Setting} & \makecell{\textbf{Job}\\\textbf{cutoff}} & \makecell{\textbf{Min}\\\textbf{interval}} & \makecell{\textbf{Exec-idle}\\\textbf{time}} & \makecell{\textbf{Exec-idle}\\\textbf{energy}} \\
\midrule
Baseline & $\geq$2 h & 5 s  & 19.17\% & 10.67\% \\
Permissive interval & $\geq$2 h & 1 s  & 23.77\% & 13.91\% \\
Conservative interval & $\geq$2 h & 10 s & 15.6\% & 7.95\% \\
Broader job set & $\geq$1 h & 5 s  & 19.22\% & 10.71\% \\
\bottomrule
\end{tabular}
\end{table}

\jpara{Sensitivity to sustained-duration threshold.}
Our baseline requires low-activity intervals to persist for at least 5\,s before counting them as execution-idle.
Under 1\,Hz passive telemetry, this is a conservative choice: a 1\,s threshold is more permissive and may include brief transients, whereas a 10\,s threshold is stricter and counts only clearly sustained intervals.
As expected, the measured magnitude changes with this threshold.
Execution-idle accounts for 23.77\% of in-execution time and 13.91\% of energy with a 1\,s threshold, compared with 15.6\% of time and 7.95\% of energy with a 10\,s threshold.
However, the qualitative conclusion is stable across all three settings.
Even under the stricter 10\,s definition, execution-idle remains substantial, showing that the phenomenon is not driven by threshold-edge events near the 5\,s cutoff.

\jpara{Sensitivity to job length.}
Our primary job-level analyses focus on jobs lasting at least 2 hours, which better capture sustained workloads while excluding short debug and interactive sessions.
This scope also matches where most energy is consumed: over the study window, jobs running at least 2 hours account for 91\% of all job-attributed GPU energy.
To test sensitivity to this filter, we repeat the analysis with a 1-hour cutoff.
As shown in~\Cref{tbl:robustness}, the resulting execution-idle estimates change negligibly, from 19.17\% to 19.22\% of in-execution time and from 10.67\% to 10.71\% of energy, indicating that our conclusions are not driven by the exact long-job threshold.

\subsection{Duration of Individual Execution-Idle Periods}
\label{subsec:temporal_structure}

Over the study window, we identify 4{,}596{,}723 execution-idle intervals. Duration of each interval matters because longer intervals keep the GPU in a loaded but low-progress state for longer, allowing elevated power draw to accumulate into more energy overhead. 

\begin{figure}[t]
    \centering
    \includegraphics[width=0.7\linewidth]{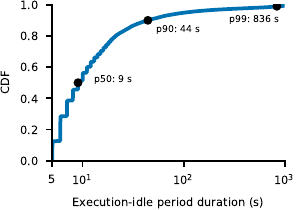}
    \caption{CDF of execution-idle interval durations.}
    \label{fig:execution_idle_duration_cdf}
\end{figure}

\jpara{Execution-idle periods can last as long as minutes.}
We conservatively require an execution-idle interval to last at least 5\,s before counting it, which already excludes short-lived fluctuations. 
In our academic-cluster measurement, even under this cutoff, \Cref{fig:execution_idle_duration_cdf} shows a median interval of 9\,s, while the 90th and 99th percentiles reach 44\,s and 836\,s, respectively.

\jpara{How much energy saving is necessary to justify a latency penalty?}
We will discuss in \S\ref{sec:implication} -- and it is well-known in the literature~\cite{frequency_scaling_time} -- that many energy savings techniques (especially frequency scaling) come with a cost in response time. 
Shifting from a low-frequency state to a high-frequency state takes some time, and hence a request arriving while the GPU is in a low energy state will suffer increased service time.
Studies report that GPUs take roughly 1--500\,ms to adjust frequency~\cite{frequency_scaling_time}.
Hence, engineers wish to avoid dropping the GPU to a low-frequency state only to immediately return to a high-frequency state: the energy savings do not justify the penalty in latency.

Reflecting on the data we observe, we philosophically wonder at what point the energy savings {\it do} justify a latency penalty for the next request. Is 44\,s (0.00267 kWh on a B200 GPU) enough? 836\,s (0.02783 kWh on a L40S GPU)?

\subsection{What other System Metrics Correlate with Execution-Idle States?}
\label{subsec:preceding_signals}

With the exception of the industry serving tests, we do not have the ability to introspect into the running code in our cluster.
As a result, it is not possible for us to causally identify what specific program behaviors are most common factors leading to execution-idle times. 
Nonetheless, we can identify probably candidates by studying what system-level conditions tend to precede execution-idle. For each execution-idle interval, we extract up to 10\,s of preceding device- and host-side telemetry, truncating the window when necessary so that it contains only the nearest preceding active-execution segment. We then apply HDBSCAN~\cite{hdbscan} to group these pre-idle windows into recurring patterns, and manually analyze the salient clusters through their telemetry signatures to assign likely causes. 

\begin{figure}[t]
    \centering
    \begin{subfigure}[t]{0.495\linewidth}
        \centering
        \includegraphics[width=\linewidth]{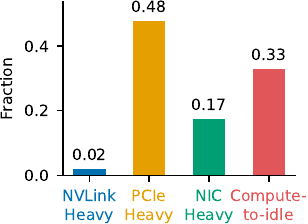}
        \caption{Labeled clusters of execution idle events}
        \label{fig:signal_category_ratio}
    \end{subfigure}
    \begin{subfigure}[t]{0.495\linewidth}
        \centering
        \includegraphics[width=\linewidth]{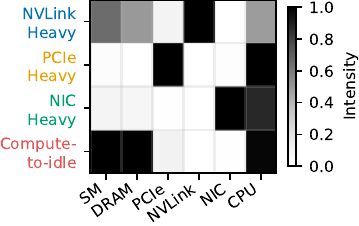}
        \caption{Signal fingerprints by cluster group}
        \label{fig:signal_fingerprints}
    \end{subfigure}
    \caption{Signals from the interval immediately preceding execution-idle, grouped into inferred categories. 
    }
    \label{fig:preceding_signals}
\end{figure}

\jpara{A small number of factors dominate the windows preceding execution-idle.}
\Cref{fig:preceding_signals} shows that most execution-idle onsets are associated with just a few categories. PCIe-heavy intervals account for the largest share at 48\%, followed by compute-to-idle at 33\%, NIC-heavy at 17\%, and NVLink-heavy at 2\%. This distribution is shaped in part by our cluster’s hardware mix. Several deployed GPU models, including L40S and 6000 Ada, do not support NVLink and therefore depend more on NIC-based communication. Accordingly, NVLink-heavy intervals are concentrated on higher-end data-center GPUs in our cluster, such as A100.

As shown in \Cref{fig:signal_fingerprints}, the fingerprints are consistent with distinct system-level behaviors. The PCIe-heavy category exhibits elevated PCIe and CPU activity, consistent with host--device transfer or coordination overhead, and is likely common in data loading, preprocessing, and framework-managed execution pipelines. The NIC-heavy category shows elevated NIC and CPU activity, suggesting distributed communication or storage-related movement such as NFS traffic, and thus aligns naturally with multi-node training and communication-intensive jobs. The NVLink-heavy category exhibits strong NVLink activity and is consistent with intra-node GPU--GPU communication in multi-GPU training on NVLink-connected servers. 

Finally, the compute-to-idle category shows elevated SM and DRAM activity immediately before idle onset, followed by a transition into near-zero activity during execution-idle. This pattern is consistent with workloads that alternate between bursts of GPU work and waiting, including bursty serving as well as training pipelines that pause at synchronization or coordination boundaries.

\subsection{Summary}
\label{subsec:findings_summary}

Across cluster-scale accounting, workload breakdowns, per-job distributions, interval durations, and pre-idle signatures, a consistent picture emerges.
Execution-idle is neither a corner case nor a purely serving-specific artifact.
It accounts for a non-trivial share of GPU energy, appears across workload classes, becomes severe for a substantial tail of jobs, often lasts long enough to matter, and tends to arise after recognizable transfer, communication, or post-compute phases.
At the same time, its cost is highly uneven, with serving workloads standing out as the most exposed regime.

\section{Implications}
\label{sec:implication}

The existence of a state in which a GPU is doing nearly zero useful work, but nonetheless consuming power intuitively demands a fix. While we will not be able to solve the problem altogether in the rest of paper, we briefly discuss three implications for developers of AI software.

\begin{myitemize}
    \item \textbf{Overall cluster utilization should not be used as a proxy metric for power draw.}(\S\ref{subsec:load_imbalance})
    \item \textbf{Ongoing research to improve utilization within a GPU are likely to improve energy efficiency.}(\S\ref{subsec:increase_utilization})
    \item \textbf{Simple, manual overrides to frequency scaling can, predictably, save energy at some latency cost.} (\S\ref{subsec:downscaling})
\end{myitemize}

\subsection{Energy Draw Can Vary Between Deployments at the Same Nominal Utilization}
\label{subsec:load_imbalance}
~\\

Operators and researchers often talk of cluster-level SM utilization figures as an implicit proxy for energy efficiency.
Because deep-idle states also consume power, operators seek to avoid leaving GPUs underutilized where they are perceived to be wasting power that could instead by applied to meaningful work.

However, energy waste due to execution-idle states means that {\it between two deployments operating at the same level of utilization, power draw can differ} and that in fact, {\it it may be better from an energy perspective to leave some cluster GPUs unutilized in favor of packing more work onto the same GPU} (at least from an energy perspective).

We perform an experiment with a biased load balancer. Rather than spreading requests evenly across the entire pool, which leaves many GPUs lightly active and repeatedly exposed to short execution-idle intervals, the scheduler  deliberately introduces \emph{load imbalance}: concentrate work onto fewer GPUs while leaving the others in deep idle state.

\jpara{Setup.}
We study this setting using an 8-GPU serving pool built by downsampling the original 96-GPU Azure Code traces~\cite{energy_tp_request_length_task_idle}.
We compare three cases: (1) a balanced baseline with all 8 GPUs active and no downscaling; (2) a 4-active-GPU case, where 4 GPUs carry the workload and the other 4 remain lightly loaded and downscaled; and (3) a 2-active-GPU case, where 2 GPUs carry the workload and the other 6 remain lightly loaded and downscaled.

\begin{figure}[t]
    \centering
    \includegraphics[width=0.7\linewidth]{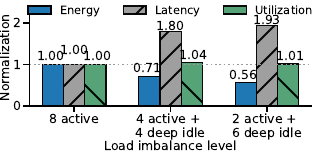}
    \caption{Energy, p95 latency, average GPU utilization at different load imbalance level, normalized to 8-active-GPU baseline.}
    \label{fig:load_imbalance}
\end{figure}

\jpara{Energy nearly halves, even though utilization stays almost the same.}
As \cref{fig:load_imbalance} shows for Azure Code, deliberately concentrating work onto fewer GPUs cuts total GPU energy to 56\% of the balanced case, while overall SM utilization changes little. Looking only at utilization would therefore suggest similar power draw across the two configurations, which is misleading.

This happens because average utilization does not reflect how energy is distributed across the GPU pool. Under imbalance, the same total work is concentrated on fewer active GPUs, while the rest remain in deep idle. Pool-wide utilization therefore stays similar, yet total energy falls because fewer GPUs remain in higher-power serving states. The energy saved by keeping more GPUs in deep idle outweighs the extra energy drawn by the more heavily loaded GPUs. In this setting, utilization masks the crucial difference between work performed and the number of GPUs still consuming baseline and execution-idle power.

\begin{algorithm}[t]
\caption{Execution-Idle-Aware Frequency Control}
\label{alg:exec_idle_control}
\small
\begin{algorithmic}[1]
\Require threshold $X$, cooldown $Y$, clocks $f_{\max}, f_{\min}$
\State $c \gets 0$, $t_{\mathrm{cooldown}} \gets 0$, $\mathit{downscaled} \gets \textbf{false}$

\For{each $\varepsilon$\,second control interval at time $t$}
    \State Read $\texttt{sm}$, $\texttt{tensor}$, $\texttt{fp16}$, $\texttt{dram}$, $\texttt{pcie}$, $\texttt{nvlink}$, $\cdots$
    \State $a_{\mathrm{comp}} \gets \max(\texttt{sm}, \texttt{tensor}, \texttt{fp16}, \cdots)$
    \State $a_{\mathrm{mem}} \gets \texttt{dram}$
    \State $a_{\mathrm{comm}} \gets \max(\texttt{pcie}, \texttt{nvlink})$

    \If{$a_{\mathrm{comp}} < 0.05$ \textbf{and} $a_{\mathrm{mem}} < 0.05$ \textbf{and} $a_{\mathrm{comm}} < 1$ GB/s}
        \State $c \gets c + \varepsilon$
    \Else
        \State $c \gets 0$
        \If{$\mathit{downscaled}$}
            \State Set GPU clock to $f_{\max}$
            \State $\mathit{downscaled} \gets \textbf{false}$
            \State $t_{\mathrm{cooldown}} \gets t + Y$
        \EndIf
    \EndIf

    \If{$c > X$ \textbf{and} $t \ge t_{\mathrm{cooldown}}$ \textbf{and} $\neg \mathit{downscaled}$}
        \State Set GPU clock to $f_{\min}$
        \State $\mathit{downscaled} \gets \textbf{true}$
    \EndIf
\EndFor
\end{algorithmic}
\end{algorithm}

\jpara{At the same time, imbalance introduces a latency penalty.}
Looking at a reduced energy cost for the same level of cluster utilization, one might be tempted to run all clusters at a deliberate skew. As load is concentrated onto fewer GPUs, serving latency rises.
With 4 active GPUs, p95 request latency increases by 80\%; with 2 active GPUs, the increases grow by 93\%, as shown in \cref{fig:load_imbalance}.
Hence, we do not prescribe highly skewed load balancing as a solution to execution idle: we merely offer it as a cautionary tale that utilization figures do not paint a complete picture of energy use.

\subsection{Increasing Utilization Within a Single GPU Does Improve Efficiency.}
\label{subsec:increase_utilization}

Although cluster-level utilization metrics are misleading, as we discuss above, individual increased GPU utilization does result in improved energy efficiency.
Because there is a fixed cost to keeping a GPU in an active and loaded state (as observed in \cref{fig:gpu_versus_cpu}, it is best to keep those GPUs busy.

Following this intuition, we suspect that \emph{co-serving systems} are likely to correlate with improved energy efficiency. These systems improve utilization by packing complementary workloads onto fewer devices, for example by co-serving online and offline jobs~\cite{qiao2025conservefinegrainedgpuharvesting}, serving multiple LLMs concurrently~\cite{yu2025prismunleashinggpusharing}, or co-serving fine-tuning and inference~\cite{oliaro2025flexllmtokenlevelcoservingllm}.

On the other hand, \emph{autoscaling systems} such as BlitzScale~\cite{autoscale3}, ServerlessLLM~\cite{autoscale2}, and INFaaS~\cite{autoscale1} have less clear energy implications. 
Their primary goal is elasticity and SLO preservation rather than direct energy minimization. By scaling in excess capacity and consolidating load, they may reduce execution-idle indirectly. However, aggressive scale-out to avoid latency degradation can also increase the number of active GPUs and thereby hurt energy efficiency. 
In this context, we see the need for direct power measurements as a key metric for system evaluation, as these systems largely report latency, throughput, utilization, or cloud cost.

\subsection{Software-induced Frequency Control is a Useful Lever Today}
\label{subsec:downscaling}

The most direct response to execution-idle is to conservatively downscale frequency during low-progress intervals.
We perform a simple experiment to override the default DVFS and instead institute our own, more aggressive algorithm in software.
Similar power-management studies and systems, such as DynamoLLM~\cite{energy_tp_request_length_task_idle} and $\mu$-Serve~\cite{serving_frequency} have used GPU frequency tuning and power capping as a configuration knob in their energy--performance optimization. 
Our goal here is narrower: we use a lightweight, local controller as a baseline to test whether execution-idle can be made less costly by reacting directly to sustained low-progress intervals.
As shown in~\Cref{alg:exec_idle_control}, our controller waits for several consecutive seconds of near-zero activity before lowering the device to the minimum available frequency, restores the original setting when activity resumes, and then holds that setting for a short cooldown period to avoid rapid oscillation.
In our implementation, the controller uses a 3\,s trigger threshold and a 5\,s cooldown period.

\begin{figure}[t]
    \includegraphics[width=0.9\linewidth]{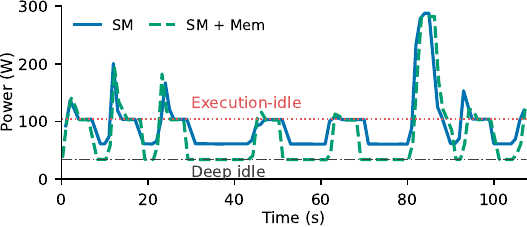}
    \caption{Power over time under SM-only and SM+memory execution-idle-aware frequency control.}
    \label{fig:frequency_downscaling_timeline}
\end{figure}

\jpara{Setup.}
We replay the Azure Code serving trace for 1175\,s on an L40S GPU under two frequency-control configurations while serving the same total number of requests.
Using \texttt{nvidia-smi}~\cite{nvidia_smi_docs}, we lower either (1) the graphics/compute clock alone or (2) both the graphics/compute and memory clocks; finer-grained component-specific controls are not exposed.
Because replay duration is fixed, we use average GPU power as a proxy for total energy.

\jpara{Downscaling more aggressively reduces the cost of deep idle states.}
\Cref{fig:frequency_downscaling_timeline} shows that online downscaling can react at the timescale of execution-idle intervals that commonly appear in our measurements (e.g., 10\,s).
On this platform, setting only the SM-related clock to the available minimum reduces execution-idle power from 105\,W to 61\,W, while lowering both SM and memory clocks further reduces it to 35\,W (deep idle power).

Across the full replay, SM-only downscaling reduces average power from 123.9\,W to 96.4\,W, a 22\% reduction.
Lowering both SM and memory clocks reduces average power further to 82.2\,W, a 34\% reduction.

\begin{figure}[t]
    \centering
    \begin{subfigure}[t]{0.495\linewidth}
        \centering
        \includegraphics[width=\linewidth]{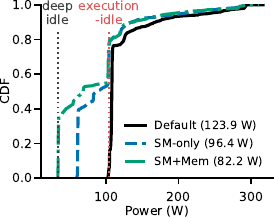}
        \caption{Power CDF (left is better)}
        \label{fig:frequency_scaling_power}
    \end{subfigure}
    \begin{subfigure}[t]{0.495\linewidth}
        \centering
        \includegraphics[width=\linewidth]{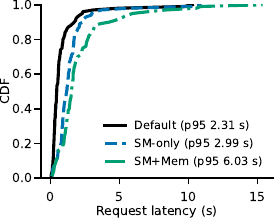}
        \caption{Latency CDF (left is better)}
        \label{fig:frequency_scaling_latency}
    \end{subfigure}
    \caption{Power--latency trade-off of execution-idle-aware frequency downscaling.}
    \label{fig:frequency_scaling_power_latency_tradeoff}
\end{figure}

\jpara{As expected, the algorithm does come with latency penalties.}
These savings, however, come with clear latency penalties.
As shown in \cref{fig:frequency_scaling_power_latency_tradeoff}, p95 latency rises from 2.31\,s to 2.99\,s (29\%) under SM-only downscaling, and to 6.03\,s (160\%) when both SM and memory clocks are reduced aggressively.

\jpara{The benefit of the approach is that it can be managed by the operator.}
Is a 160\% increase in latency an acceptable cost in exchange for a 34\% power reduction? Is a 29\% increase in latency in exchange for a 22\% reduction a better deal? Or is no latency cost acceptable?

From a hardware vendor's perspective, GPUs typically compete on performance and so it is no surprise that default GPU configurations opt for better latency and higher energy. However, for operators and developers who may wish to strike a different bargain, the levers already exist to make a different choice (and likely, with application-aware context, an even better choice than the simple and na\"{i}ve algorithm above).

\

\section{What Broader System Support is Needed to Manage Execution-Idle in Future Systems?}
\label{sec:future}

Before concluding, we reflect on a few research directions that might improve energy waste due to execution-idle periods.

\jpara{Workload--power interfaces for execution-idle management.}
Execution-idle should not be treated purely as a device-level phenomenon.
Its frequency, duration, and performance sensitivity depend strongly on workload structure: some jobs can tolerate aggressive downscaling during low-progress periods, while others cannot.
This suggests a workload--power co-design opportunity, where applications, system software, or serving frameworks expose signals such as burstiness, slack, communication phases, or latency sensitivity to lower layers.
With this information, the system could better decide when execution-idle is safe to exploit, when it lies on the critical path, and how aggressively to trade performance for energy.
More generally, systems should expose the power implications of execution-idle explicitly, rather than forcing hardware control to infer them indirectly from utilization counters alone.

\jpara{SLO-aware execution-idle control for latency-sensitive serving.}
For latency-sensitive services, execution-idle-aware control should be integrated into SLO-driven resource management rather than applied in isolation.
Prior work such as \mbox{$\mu$-Serve}~\cite{serving_frequency} and DynamoLLM~\cite{energy_tp_request_length_task_idle} shows that dynamic GPU-frequency control can reduce serving energy while meeting latency SLOs.
Our results suggest that execution-idle provides a complementary signal for such policies: even under SLO-aware control, it can still arise as a distinct low-progress regime within execution, where deeper temporary downscaling may be worthwhile.
An important open question is how to combine execution-idle detection with queueing state, burst forecasts, slack, and tail-latency objectives in a unified serving controller.

\jpara{From device-level to component-aware power proportionality.}
Our study focuses on device-level power proportionality: execution-idle keeps whole-device GPU power elevated despite near-zero visible activity, and current controls can reduce part of that cost.
At the same time, our results suggest that this inefficiency is not tied to a single component: lowering SM frequency reduces execution-idle power substantially, and lowering memory frequency reduces it further.
A natural next step is component-aware power proportionality.
Future systems could ask whether execution-idle also exists within individual subsystems, such as compute, memory, and communication, and whether components off the critical path can be downscaled independently.
Realizing this would require richer component-level observability and more flexible control across GPU subsystems.

In summary, these directions reinforce the same message: execution-idle should be treated as an important power state.
It is a recurring and costly operating regime, and future systems should detect it, reason about it, and manage it explicitly.
\section{Related Work}
\label{sec:related}

\jpara{Energy proportionality has been studied extensively in CPU-centric systems.}
Classic work~\cite{cpu_energy1,cpu_energy2} on energy-proportional computing asks how closely server or CPU power tracks utilization.
This question is now especially important for GPUs.
As AI demand grows, GPUs account for an increasing share of data-center power~\cite{epri2026powering_intelligence,shehabi2024datacenterenergy,mckinsey2024ai_power}, making GPU energy efficiency increasingly critical.
As a result, GPU power proportionality is no longer merely a device-level concern, but an important determinant of overall data-center energy efficiency.

\jpara{Prior GPU energy work focuses mainly on aggregate efficiency and operating points.}
A broad body of important work takes an end-to-end perspective on GPU energy, studying metrics such as energy per token or energy to solution~\cite{bloom}, and evaluating how efficiency varies with frequency~\cite{energy_frequency1,energy_frequency2,energy_frequency3}, batch size~\cite{energy_tokennum_batch,energy_batch_models,energy_batchsize_limit}, request shape~\cite{energy_tokennum_batch,energy_token_num}, model choice~\cite{energy_batch_models,energy_model_task_gpu,luccioni_task_power}, serving engine~\cite{engine_benchmark}, hardware platform~\cite{ml_energy,energy_model_task_gpu}, and request load~\cite{energy_rps1,energy_rps2}.
This literature has been invaluable in establishing that GPU energy efficiency is highly workload- and configuration-dependent.

\jpara{Recent measurement studies characterize GPU power and utilization across phases and deployments.}
Recent work measures power-management opportunities for LLM inference in the cloud and characterizes GPU utilization patterns in large-scale systems~\cite{characterize_llm_power,single_node_measurement,energy_rps1,energy_batch_models,inference_measurement1,amd_profiling}.
These studies show substantial variation in GPU power and activity across jobs, phases, and deployment settings, and they provide important foundations for reasoning about GPU energy beyond coarse aggregate metrics.

\jpara{Runtime power-management and serving systems provide important control mechanisms.}
Related serving and control systems such as $\mu$-Serve~\cite{serving_frequency}, DynamoLLM~\cite{energy_tp_request_length_task_idle}, BlitzScale~\cite{autoscale3}, ServerlessLLM~\cite{autoscale2} and vendor mechanisms such as Blackwell datacenter power profiles~\cite{blackwell_power_profiles,blackwell_power_profiles_paper} show how energy can be improved through frequency control, autoscaling, and device-level reconfiguration.
These mechanisms are closely related to our setting because they can change either the cost of low-activity periods or the system conditions under which such periods arise.

\jpara{Taken together, this literature shows that understanding GPU energy requires looking beyond aggregate utilization to how power and activity co-evolve over time, and to where GPU energy is spent productively versus unproductively.}
Our work complements these efforts by isolating \emph{execution-idle} as a recurring loaded-but-low-activity regime, quantifying its prevalence and cost in real deployments, and showing why it warrants explicit attention in future GPU system design.

\section{Conclusion}
\label{sec:conclusion}

We identify \emph{execution-idle} as a distinct GPU operating regime in which a program remains loaded, visible activity is near zero, yet power remains well above deep idle. Across a 31-day cluster study and replayed serving traces, we show that execution-idle is common, consumes a meaningful share of GPU energy, and is especially important for bursty serving workloads. We hypothesize that this cost arises because current GPU power behavior is tuned to preserve responsiveness across brief stalls by keeping clocks and power elevated; in real workloads, however, these low-activity periods often persist long enough for the energy cost to accumulate. We further show that execution-idle can be mitigated either by lowering its cost through downscaling or by reducing exposure to it through scheduling, although both approaches introduce explicit energy--performance trade-offs. Taken together, these findings argue that future GPU systems should treat execution-idle as a first-class operating state and manage it explicitly in the pursuit of more energy-efficient AI infrastructure.

\bibliographystyle{ACM-Reference-Format}
\bibliography{paper}

\appendix \section{Compute Infrastructure and Cluster Specifications}
\label{app:cluster-specs}

The measurements in this paper were collected on a large academic AI/HPC cluster.
The cluster contains a heterogeneous NVIDIA GPU fleet spanning Ampere, Ada Lovelace, Hopper, and Blackwell generations.
Importantly, the \emph{installed} fleet is larger than the subset used in our telemetry study: not all GPUs could be included in the final dataset because some device generations had compatibility issues with our profiling stack (e.g., incomplete or unstable support in the telemetry tools used for continuous collection).
Our analysis therefore uses the subset of devices for which profiling was reliable throughout the study period.

\subsection{System Overview and Hardware Configuration}

Table~\ref{tab:cluster-overview} summarizes the host environment, interconnect, and software stack.
The cluster comprises 8{,}288 CPU cores and an installed fleet of 880 GPUs.
Nodes run a production software stack based on AlmaLinux, Slurm, recent NVIDIA drivers, CUDA, and DCGM.

\begin{table}[t]
\centering
\small
\renewcommand{\arraystretch}{1.1}
\begin{tabular}{@{}ll@{}}
\toprule
\textbf{CPU Cores} & 8{,}288 \\
\textbf{Installed GPUs} & 880 NVIDIA GPUs \\
\textbf{GPU Generations} & Ampere, Ada Lovelace, Hopper, Blackwell \\
\textbf{Interconnect} & 100\,Gbps EDR InfiniBand (compute \& storage) \\
                         & 1\,GbE Ethernet (node management) \\
                         & 10\,GbE Ethernet (external routing) \\
\textbf{OS}              & AlmaLinux 9.5 (Teal Serval) \\
\textbf{Kernel}          & 5.14.0-503.40.1.el9\_5.x86\_64 \\
\textbf{Job Scheduler}   & Slurm 24.05.0 with QoS-aware preemption \\
\textbf{NVIDIA Stack}    & Driver 575.51.03, CUDA 12.9, DCGM 3.3.9 \\
\bottomrule
\end{tabular}
\caption{Overview of the academic GPU cluster used in this study.}
\label{tab:cluster-overview}
\end{table}

\subsection{GPU Fleet Composition}

Table~\ref{tab:gpu-fleet} shows the installed GPU fleet and the corresponding default driver-enforced power limits reported by \texttt{nvidia-smi}.
These counts describe the installed cluster hardware; as noted above, the profiled subset is smaller because telemetry collection was not uniformly supported across all device/tool combinations.

\begin{table}[t]
\centering
\small
\renewcommand{\arraystretch}{1.1}
\begin{tabular}{lrr}
\toprule
\textbf{GPU Model} & \textbf{Count} & \textbf{Set Power Limit} \\
\midrule
L40S                      & 410 & 400\,W \\
RTX A6000                 & 208 & 300\,W \\
RTX 6000 Ada Generation   & 58  & 300\,W \\
L40                       & 56  & 300\,W \\
A100 80GB (PCIe)          & 48  & 300\,W \\
RTX PRO 6000 (Blackwell)  & 40  & 600\,W \\
A100 40GB (SXM4)          & 24  & 400\,W \\
H100 (SXM5)               & 24  & 700\,W \\
B200                      & 8   & 1000\,W \\
H200 (SXM)                & 4   & 700\,W \\
\midrule
\textbf{Total}            & \textbf{880} & \textit{Persistence Mode Enabled} \\
\bottomrule
\end{tabular}
\caption{Installed GPU fleet composition and default power limits reported by \texttt{nvidia-smi}.}
\label{tab:gpu-fleet}
\end{table}

\subsection{Network Topology}

The cluster uses a dual-network design that separates management traffic from high-throughput data transfer.
Compute nodes are equipped with NVIDIA/Mellanox MT4123 InfiniBand host channel adapters operating at 100\,Gb/s for inter-node communication and storage access.
Standard node management uses 1\,GbE, while external routing uses 10\,GbE.
GPUDirect RDMA is not used for inter-GPU communication in this deployment.

\subsection{GPU Configuration and Profiling Coverage}

All GPUs operate in their factory-default configuration.
NVIDIA Persistence Mode~\cite{nvidia_persistence_mode} is enabled, while Applications Clocks are left inactive.
The driver therefore manages runtime DVFS and GPU Boost subject to the default hardware power limits, without manual clock locking or job-specific power caps.

\end{document}